\documentclass[a4paper,12pt]{article}
\usepackage{color}
\usepackage[a4paper,colorlinks=true,urlcolor=blue,citecolor=blue,linkcolor=blue,bookmarks=false]{hyperref}
\usepackage[pdftex]{graphicx}
\DeclareGraphicsExtensions{.pdf,.png,.jpg,.gif}
\usepackage{xspace}
\usepackage[top=2cm,bottom=2cm,left=2cm,right=2cm]{geometry}
\usepackage[latin1]{inputenc}

\usepackage{listings}

\usepackage{enumitem}
\def\myfig#1{Fig.~#1\xspace}

\begin{document}

\title{Cryptographically verifiable anonymous voting using pan-european e-id's
\\
{\normalsize An analysis of the STORK 2.0 framework}
}
\author{\normalsize Alessandro Preziosi (alessandro.preziosi@polito.it)
\\
\normalsize Diana Berbecaru (diana.berbecaru@polito.it)}
\date{November 28, 2016}
\maketitle

\vfill

\rule{\textwidth}{1pt}

\newpage
\tableofcontents

\rule{\textwidth}{1pt}

\vfill

\newpage

\section{Introduction}

The STORK 2.0 project (Secure idenTity acrOss boRders linKed 2.0) aims to realize a single European electronic identification and authentication area.  The purpose of the project is to facilitate the citizens' lives and to improve citizen and business mobility inside the European Union. The interoperability of eID solutions has many practical applications. For example, a German student studying in Spain could identify to the Spanish university using his account at the German university, without having to verify his identity in person, create a second account, and store all the necessary data in two different places. 

In the future, this framework could allow people to start a company, get their tax refund, or obtain their university papers without physical presence.

The role of the STORK platform is to identify a user who is in a session with a service provider, and to send his data to this service. Whilst the service provider may request various data items, the user will always controls the data to be sent. 

The different agents involved in the authentication process are:

\begin{itemize}
   \item The \textbf{Service Provider} (SP), which offers a service to the end user;
   \item The \textbf{Pan European Proxy Service} (PEPS). One for each country, used to proxy requests across countries;
   \item The \textbf{Identity Provider} (IdP), to which the  user can authenticate;
   \item The \textbf{Attribute Providers} (AP), supplying additional information about the user;
\end{itemize}

For a more detailed explanation of the STORK architecture see \cite{stork1} and \cite{stork2}.

In this work we will focus on the possibility of creating anonymous services on top of the STORK architecture. This can be useful in particular in the case of electronic surveys or elections. In this case we want the users to be authenticated, to prevent repeated voting, but at the same time the votes should be anonymous. It should be impossible to find out how a specific person voted. 

In order to do this blind signatures and an onion routing system similar to the one used in TOR\cite{tor} (\url{https://www.torproject.org/}) are implemented. 

Source code for the STORK 2.0 framework, used in all the following sections, can be found at \cite{storkSource}

The purpose of this work is to describe this anonymity protocol in detail, to anlyze a reference implementation, and highlight potential weaknesses.  

\newpage

\section{Protocol description}

To describe the protocol we will use an analogy with real life voting. In our case the Ballot Box is a server owned by the service provider, and the voting tickets are packages sent through the network.

In order to make it impossible to identify voters, the packages are bounced through a number of proxy nodes before reaching the ballot box, making it as hard as possible to trace their origin. We cannot allow anyone to vote as many times as they want, therefore tickets must be signed, but the service provider cannot sign the plaintext tickets, otherwise anonymity would be lost. Therefore a blind signature scheme is used, to provide \textit{unlinkability}.

\subsection{Blind signature}

A \textit{blind signature} is a system through which an entity can sign a message with no knowledge of the message itself.

In this case the RSA blind signature scheme described by Chaum is implemented\cite{chaum}.

\begin{itemize}
   \item First, the client generates a big random number \textit{r}, and raises it to the public exponent, modulo N. This is called the \textit{blinding factor}.
   \item The message is then multiplied by the \textit{blinding factor}, and the result is sent to the SP for signing.
   \item The SP signs the blinded message and returns the signature \textit{s} to the client.
\end{itemize}

Due to the mathematical properties of RSA signatures, if the client divides \textit{s} by \textit{r} modulo N, he will obtain a valid signature of the original message. For more details see \cite{chaum}.

In order for the blind signature to be secure the number \textit{r} should be big and truly random, otherwise it can be brute forced by the signing entity. Let's check if this is true for the reference implementation. 

\subsubsection{Brute forcing the opacity factor}

In the demo-client.js the random number is generated by the function \verb|rndInt32()|.
The size of the opacity factor used by the demo client is 64 bits: 

(demo-client.js:627)
\begin{lstlisting}[frame=single]
opc[I]=k=new eSvyBigInteger(rndG.rndInt32()+''+rndG.rndInt32(),10);
\end{lstlisting}

A length of 64 bits could be considered secure in this use case, but a longer opacity factor could be preferred. With a distributed brute force attack it could be possible for the SP to unblind the ticket of a specific voter, and find out who he voted for. This would take a couple years with current technology, in order to unblind a single ticket (in 2002, the Distributed.net project cracked a 64-bit RC5 key in 1,757 days using at total of 331,252 computers\cite{distributed}).

\label{entropyWeakness}
But this specific implementation is much weaker for another reason. In fact, it is not sufficient for the opacity factor to be big enough, but it must also be truly random. 

The client uses the function \verb|eSvyPrng()| as an entropy buffer.

All the data used for the buffer is basically a fingerprint of the browser (resolution, platform, available plugins, etc.), with the addition of two calls to Math.random(). 

If we look at the Math.random() implementation in major browsers, it is not cryptographically secure. In fact it is usually a pseudo-random generator seeded with the time of the first call in milliseconds. For details on the different implementations see \cite{random}.

This means that if an attacker can run javascript on the voter's client, and he knows the time when the opacity factor was generated, he can easily brute force the opacity factor. 

That information is always available to the SP, but it can also be available to a third party attacker who could obtain the timing information by sniffing the connection between the client and the SP as well, and the browser fingerprint by managing to make the voter run some javascript in his browser. 

The real entropy of the opacity factor is thus reduced to just the time of generation of the pseudo random number, in milliseconds. This can easily be known with little uncertainty, thus brute forcing the opacity factor would be trivial. 

In order to prevent this, a better random generator with an improved entropy pool should be used. For example, Fortuna\cite{fortuna} is a javascript crypto library with a strong pseudo number implementation which accumulates entropy from various sources including:

\begin{itemize}[noitemsep,nolistsep]
  \item mouse movements
  \item internal clock
  \item keyboard activity
  \item the built-in crypto.getRandomValues() of some modern browsers
\end{itemize}

\subsubsection{The One-More-RSA-Inversion Problem}

In a blind signature scheme it is also desired that signing one blinded message produces at most one valid signed messages. In the case of Chaum's scheme case this is not true\cite{chaumsec}. In particular, the blinded message and the unblinded signature \textit{s} would also constitute a correct combination, which could potentially allow a user to vote twice. This can be avoided by asking users to sign a hash of the message instead of the message itself. 

In our the ticket is signed directly, but this is only a security concern if given the structure of the ticket, it is possible to easily find a blinding factor which, multiplied by the original ticket (modulo N) still generates a valid ticket. Given the relative complexity of a valid ticket in this case, I would argue it would be computationally unfeasible to craft such an opacity factor and use it to double vote. 

\subsubsection{RSA blinding attack}

It is also important to be aware of the \textit{RSA blinding attack} when using this blind signature scheme. With this attack the signer can be tricked into decrypting a message by blind signing another message.  

This happens because the signing process is equivalent to decrypting with the signer's private key. An attacker can thus provide a blinded version of an encrypted message for the SP to sign, in order to decrypt the message.
It is therefore important that the SP does not use the same key for signing tickets and for encryption.

The SP implementation in this case doesn't suffer from this, as a new, survey-specific, private key is generated for each survey:

\newpage
(survey.php:225)
\begin{lstlisting}[frame=single,breaklines=true,basicstyle=\footnotesize]
function generateNewSurvey(){
  global $keyLength;
  list($certificate,$key,$module,$exponent)=newKeyandCert($keyLength);
  mysql_query("insert into Surveys (privatekey,module,exponent,certificate) ".
              "values ('$key','$module','$exponent','$certificate')"); 
  return mysql_insert_id();
}
\end{lstlisting}

(wserv.php:255)
\begin{lstlisting}[frame=single,breaklines=true,basicstyle=\footnotesize]
  //Signature MUST BE survey-specific, to avoid cross-survey
  //authorizations (and thus, multiple participations for a single
  //citizen)
  if (!$surveyId=$_REQUEST['idsvy'])
    errorDie(999,'You must provide a survey id');
  //Get the survey-specific signing key
  list($sKeyS)=mysql_fetch_row(mysql_query("select privatekey from Surveys where surveyId = '$surveyId'"));
  $k=openssl_pkey_get_private($sKeyS);
\end{lstlisting}

\subsection{Onion routing}

In order to make voting anonymous it is not sufficient to use a blinding signature scheme. In fact, if the signed tickets were then sent directly by the voters to the SP, it would be easy to deanonymize a voter by looking at the time the vote was sent and/or the sender IP address. 
Therefore it is necessary to hide this information in one way or another. 

An anonymization network like the one used in TOR\cite{tor} could be useful to hide the voter's IP, and make it impossible for the SP to know which IP address the ticket was sent from. In TOR every package is proxied through several routers before reaching its destination. Packages are encrypted several times, so they have with multiple layers, like an onion. Each layer is encrypted with the public key of a specific node in the route, so that every node will only be able to read who's next in the route, but not the contents of the packet or its real destination\cite{tor}.

This is used to hide the sender IP address, but in this case it's not enough. In fact, by knowing the time when the ticket was blindly signed, the SP could easily identify the user who sent a certain ticket by looking at the time the ticket was received. This kind of time-based linking will not be 100\% accurate, but it can be extremely accurate if the voters are not too many. 

It is therefore paramount to make the votes reach the Ballot Box in a truly random order, so they cannot be linked with the user who sent them. 

In order to achieve this, in the STORK anonymity network, packages are not relayed as fast as possible, but according to a specific delay set by the user who created the package.

The SP will choose an interval during which all tickets for a specific survey must be collected. EndDate (\textit{endD}) is the time when no more participations can be issued, and the ballot box will start accepting tickets. CloseDate (\textit{cloD}) is the time when the ticket collection ends. 

It will be the client to decide at which time his packets will be relayed between the nodes in the network, and the time of final delivery to the ballot box. 

These timings are picked backwards, starting from the time of delivery to the ballot box.

\begin{figure}[!h]
\centerline{\includegraphics[width=0.7\textwidth]{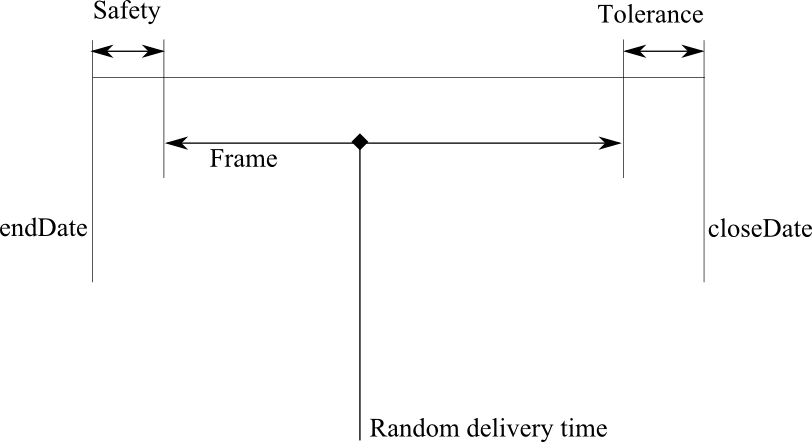}}
\caption{The delivery time is picked at random within the frame}
\label{fig:frame}
\end{figure}

The random delivery time is picked at random within a "frame", which starts at \textit{endD + safety} and ends at \textit{cloD - tolerance}, as shown in \myfig{\ref{fig:frame}}. Tolerance is a value to avoid packets reaching the ballot box too late. It is hardcoded (2 minutes in this case), and the choice depends mainly on how often the nodes in the network check their queue to forward packets (which is roughly the upper limit of the maximum delay a package should experience). 

The \textit{safety} parameter is a time, in seconds, which represents the minimum acceptable delay, to avoid packages going through the anonymity network with no delay at all.

\begin{figure}[!h]
\centerline{\includegraphics[width=0.7\textwidth]{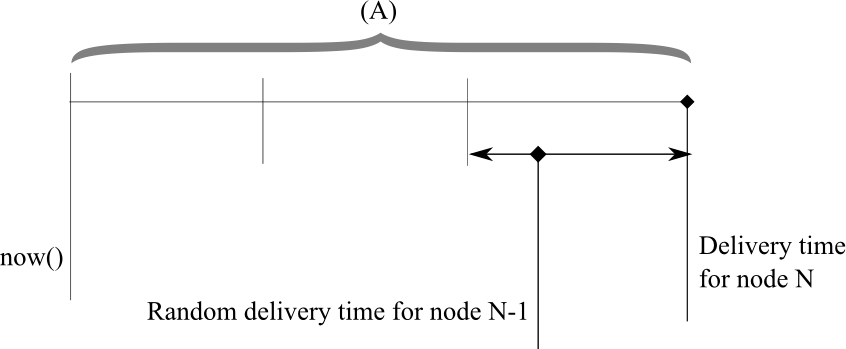}}
\caption{The remaining time (A) is divided by the number of remaining nodes (3 in this case) to pick a random delivery time for each node.}
\label{fig:frame2}
\end{figure}

The client will pick a random delivery time to send the packet to the ballot box within this frame. Then all the delivery times for the previous steps in the onion routing will be picked, starting from the last node and going back, as shown in \myfig{\ref{fig:frame2}}.

For example, let's say \textit{endD}=1000, \textit{cloD}=1200 and \verb|now()| = 100. A randomly chosen delivery time for the ballot box could be 1100. Then the remaining available time is calculated, as 1100-\verb|now()| = 1000, then divided by the number of remaining nodes on the route, let's assume 10. The delivery time for node number 9 would be picked at random between 1100 (the time of delivery to the ballot box) and (1100 - (1000/10)). This calculation is repeated recursively for all remaining nodes on the route, from the last one to the first, to choose the delivery time for every step in the route.

This should provide untraceability of the voters, as all votes are received by the ballot box at a random random between the closing time and the end time, from random nodes in the network. 

\label{cloDattack}
The problem is that a malicious SP could provide different clients with different values of \textit{cloD} and \textit{endD}, thus identifying voters by the time their ticket is received. 

Therefore it is important that all voters are served the same values of cloD and endD. This must be verified before the packets are sent into the network. To do this, a hash of these parameters could be published on some third party server. 

\subsection{The protocol in detail}

\subsubsection{Authentication}

Before being allowed to participate every user must be authenticated to the SP. This authentication can happen through any means and must not necessarily use the STORK network. 

The authentication is needed to avoid multiple voting or voting by people who shouldn't be allowed to take part in the survey. It also lowers the surface area for a Denial Of Service attack, as any unauthenticated request to the SP will be immediately dropped.  

In the case of the demo-SP, authentication is not implemented. A dummy function is used, which always returns true. Every user who connects to the demo SP is considered as authenticated and is given access to the survey.

(wserv.php:400)
\begin{lstlisting}[frame=single,breaklines=true,basicstyle=\footnotesize]
function checkAuthenticationAndAuthorization($surveyId='',$mark=false) {
  // Here, being authenticated and authorised should be checked (to
  // participate if survey id is given, and to use the proxy if
  // not). Always true on the demoSP
  if ($mark)
    // Here participant should be marked so he cannot participate twice.
    return true;
  return true;
}
\end{lstlisting}

\subsubsection{Client initialization}

\begin{figure}[!h]
\centerline{\includegraphics[width=0.7\textwidth]{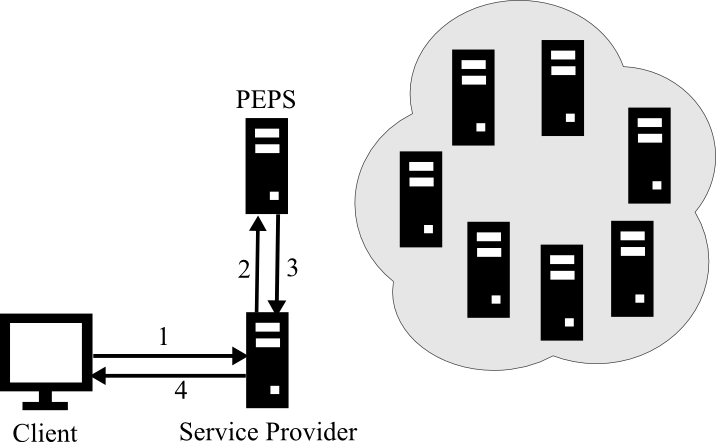}}
\caption{The information about the survey and the list of nodes are retrieved from the SP through https}
\label{fig:stage1}
\end{figure}

Initially the client needs to download the list of nodes in the network. This list is retrieved from the S-PEPS. 

(common.php:73)
\begin{lstlisting}[frame=single,breaklines=true,basicstyle=\footnotesize]
$nodelistDownloadURL=$SPEPSHost.'/PEPS/anonymity/resources/nodes/citizen/js';
\end{lstlisting}

To avoid problems with ssl certificates, in the demo the list of nodes is not taken directly from the PEPS but from the SP (\myfig{\ref{fig:stage1}}, step 1) which forwards the request to the PEPS (step 2), then sends it back to the client as a javascript, through https (step 4): 

(survey.php:184). 
\begin{lstlisting}[frame=single,breaklines=true,basicstyle=\footnotesize]
if($TESTMODE){
    //Client expects the node list from a secure url. demoSP gets it and
    //delivers to the client through an https connection (Client will
    //already be trusting demoSP's certificatem eventhough it is self
    //signed)
    $nodelistURL='?op=nodesjs';
}
\end{lstlisting}

The demo client always checks that all data is loaded through https:

(demo-client.js:134)
\begin{lstlisting}[frame=single,breaklines=true,basicstyle=\footnotesize]
for (var i in scripts){
    if(scripts[i].match(/^(http?|HTTP?)/) && !scripts[i].match(/^(https:?|HTTPS:?)/)){
        alert(_('Safety threat: Critical data was retrieved through an unsecured channel.\n\n'+scripts[i]+'\n\nAborting.'));
        stat='err';
        this.ended = true;
        break;
    }
}
\end{lstlisting}

This is true in this case as well, because the survey page will be delivered through https, and the javascript client will trust the survey's certificate even if it's self signed (because it's on the same page). 

On a deployment configuration the client should download this list directly from the PEPS, through a secure SSL connection.
This is important otherwise an attacker could serve a list of nodes controlled by him. This would not enough to deanonymize voters though, unless the attacker also has access to the ballot box server.

The list is received as a javascript script which populates the variable \verb|nodes|. 

If the client is not javascript and not limited to the same-origin policy, the list can also be retreived in JSON or XML format (services/ResourcesService.java), from the endpoints:

\begin{verbatim}
/PEPS/anonymity/resources/nodes/citizen/json
/PEPS/anonymity/resources/nodes/citizen/xml
\end{verbatim}

For every node in the list the client receives its url, its public key, and a weight.

The weight is used for load balancing when picking a route at random, so nodes with a higher weight get picked with a higher probability. 

The client also needs information regarding the survey and the SP. This is passed through a variable \verb|parameters| generated by survey.php.

These are the elements contained in the anonymousSurvey XML structure:

\begin{itemize}
\item \textbf{name} : name of the survey (only chars and numbers)
\item \textbf{idSvy} : id of the survey, generated by the SP (only numbers)
\item \textbf{lang} : language to be used for client messages, two letter ISO code
\item \textbf{srvUrl} : URL of the Service Provider
\item \textbf{srvAuth} : token for authentication and authorization towards SP. Will be sent along with each connection and SP must reject them if the token is not valid (user is no longer authenticated or authorised)
\item \textbf{srvCert} : survey specific RSA key modulus for blind signature (base 64 encoded)
\item \textbf{SRVeXP} : public exponent for previous modulus (base 64 encoded)
\item \textbf{rouLen} : route length (recommended 3 ... 6), 0 = direct to ballot box through SP proxy
\item \textbf{bBxUrl} : absolute URL of the ballot box
\item \textbf{bBxMod} : public key (modulus) of the ballot box (base 64 encoded)
\item \textbf{bBxExp} : public key (exponent) of the ballot box (base 64 encoded)
\item \textbf{endD} : datetime (ISO string or timestamp in miliseconds) of the end of polling (when no more participations can be issued)
\item \textbf{cloD} : datetime (ISO string or timestamp in miliseconds) when the tallying will start (at least cloD-endD \textgreater= 600+rouLen*120, if 0, the client will assume endD+2(600+rouLen*120)
\item \textbf{areaLog} : name attribute value of a textarea to show logs to the user, null if not available 
\item \textbf{disExt} : true to disable the automatic launch of browser extension
\item \textbf{urlExt} : URL for an informational page (for the browser extension)
\item \textbf{pause} : optional: true to start eSurvey paused (routes won't be calculated before hitting submit). Else, route calculation will start on page load
\item \textbf{pauseIfExt} : optional: true to start eSurvey paused when the extension is present, call the iterate method to start
\item \textbf{refresh} : true if the extension must invite the user to reload the page after submitting the participation
\item \textbf{minVer} : minimum required version of the client software
\item \textbf{keepA} : To send periodic connections to SP to keep the session alive while the survey is being filled in
\item \textbf{skip} : name of an element of the form to be skipped (will not be sent)
\end{itemize}

An example survey would look like this:

\begin{verbatim}
<anonymousSurvey>                         
   <name>surveyForm</name>                             
   <idSvy>845</idSvy>                               
   <lang>en</lang>                               
   <svrUrl>wserv.php</svrUrl>              
   <svrAuth></svrAuth>                     
 <svrSCert>9eF5Lm7iEncZDxE...mpw241Z8gbfcOjPVropK845M=</svrSCert>
   <svrSExp>AQAB</svrSExp>              
   <keepA>true</keepA>                           
   <rouLen>2</rouLen>                    
   <bBxUrl>https://.../demoSP/wserv.php</bBxUrl>    
   <bBxMod>AMhgf21MMHKYJY...9rWbFrC55K0iNn8xjaQPJrn</bBxMod>
   <bBxExp>AQAB</bBxExp>                      
   <endD>1395820821000</endD>                   
   <cloD>1395822501000</cloD>            
   <sButt>sendButton</sButt>                           
   <areaLog>logarea</areaLog>                       
   <pause>true</pause>                     
   <disExt>true</disExt>                   
</anonymousSurvey>                              
\end{verbatim}

(The srvAuth is always empty in the demoSP code, but it should be a valid authentication token)

Before starting the voting process the client asks the SP if the user is authenticated by sending a request containing the surveyId and the authentication token.

Example request: \verb|param=auth&token=...&idsvy=999|

Example response: \verb|<response><authorised/></response>|

In the case of the demoSP the answer is always positive.

The client will also ask the SP the current time to avoid problems due to date sincronization when calculating delivery times.

Example request: \verb|param=date&token=...&idsvy=999|

Example response: 

\verb|<response><serverTime format="unix">1395820534</serverTime></response>|

\newpage
\subsubsection{Checking that nodes are alive}

\begin{figure}[!h]
\centerline{\includegraphics[width=0.7\textwidth]{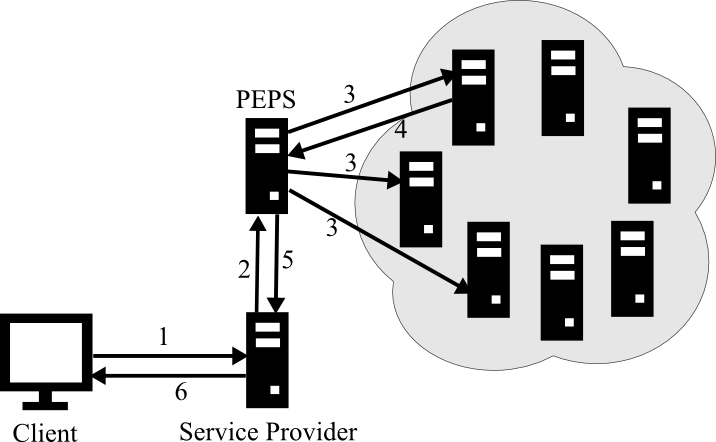}}
\caption{A bundle containing challenges for the first node of each route is signed by the SP and then relayed to the network by the PEPS, to check which nodes are alive. The response is proxied back to the client}
\label{fig:stage2}
\end{figure}

At this point the client generates the random routes the ticket will take. 

For each alternative route, the first node in the route is challenged to check if it's up and working. 

A challenge package is like a participation package but with only one layer of encryption, and no relay data, only the challenge. The nodes will just decipher it, send back the challenge, and discard the package. 
Packages are encapsulated inside a \textit{bundle}, which is sent to the SP (\myfig{\ref{fig:stage2}}, step 1).
The request to proxy the package to the peps looks like this:

\begin{verbatim}
param=proxy
token=dl0tim56ujtlu2pd2509cdm7j2
idsvy=845
bundle=
<bundle>
 <sealedPackage>
   <id>0</id>
   <payload>...</payload>
   <key>...</key>
   <recipient>.../PEPS/anonymity/handler/node</recipient>
 </sealedPackage>
</bundle>
\end{verbatim}

The SP will sign the bundle with XMLdsig and then relay it to the PEPS (\myfig{\ref{fig:stage2}}, step 2): 

(wserv.php:300) 
\begin{lstlisting}[frame=single,breaklines=true,basicstyle=\footnotesize]
//We sign the bundle (or else, it won't be accepted by the S-PEPS)
$bundle = calculateXMLDsig($bundle,$demoSPCertFile,$demoSPKeyFile);
\end{lstlisting}

The PEPS will check the signature and if it's not a valid signature from a trusted SP (or node), the package will be discarded. This allows only trusted entities to forward packets into the network to prevent some Denial of Service attacks. 

(InboundPackageHandler.java:330)
\begin{lstlisting}[frame=single,breaklines=true,basicstyle=\footnotesize]
Bundle bundle = Bundle.getBundleFromXMLString(strBundle);

checkBundleSignature(strBundle, NO_PROXY);
LOG.trace("Bundle signature check!");
\end{lstlisting}

If the signature is correct, the PEPS will send the packages inside the bundle to the respective nodes (\myfig{\ref{fig:stage2}}, step 3).
When the PEPS wil receive the responses from the nodes (\myfig{\ref{fig:stage2}}, step 4) it will relay them to the SP (\myfig{\ref{fig:stage2}}, step 5), which will again relay them to the client (\myfig{\ref{fig:stage2}}, step 6).

\begin{figure}[p]
\centerline{\includegraphics[width=1.0\textwidth]{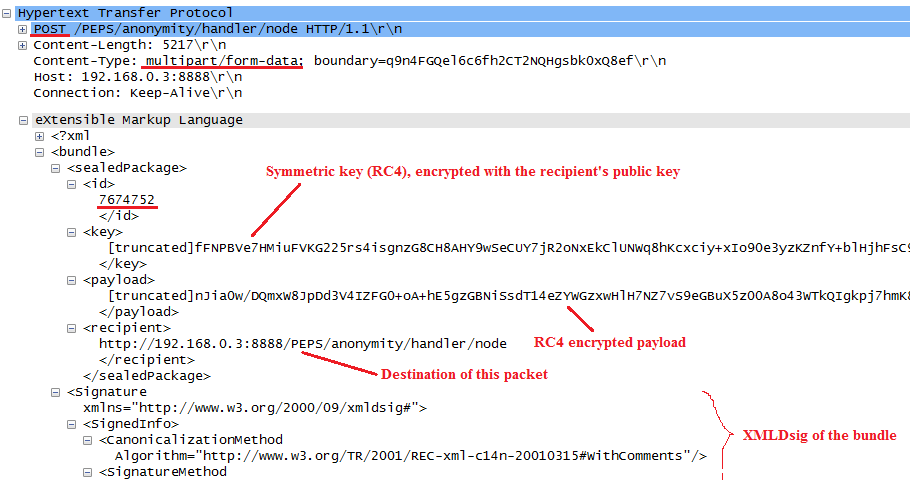}}
\caption{Request}
\label{fig:request}
\end{figure}

\begin{figure}
\centerline{\includegraphics[width=1.0\textwidth]{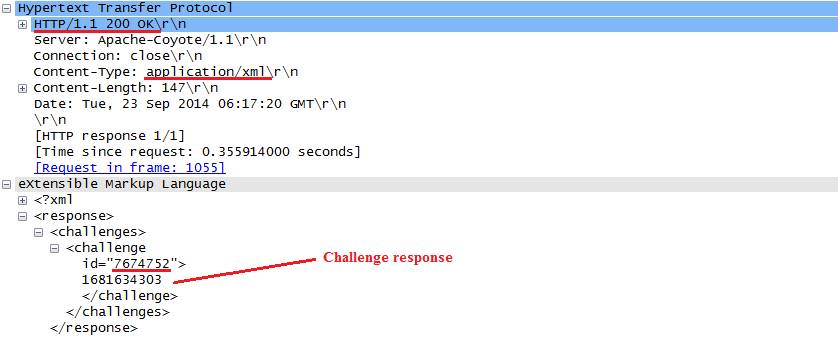}}
\caption{Response}
\label{fig:response}
\end{figure}

An example of request with a signed bundle can be seen in \myfig{\ref{fig:request}}. \myfig{\ref{fig:response}} shows the response from the node. 

If the client doesn't receive any challenge response from any of the first nodes in the alternative routes, the process is interrupted and he will have to try again a later time. 

\subsubsection{Package creation}

At this point the client generates a ticket with this structure:

\begin{verbatim}
<ticket>
	<idTicket>12345</idTicket>
	<idSurvey>12345</idSurvey>
    <endTime>endD</endTime>
	<closeTime>cloD</closeTime>
</ticket>
\end{verbatim}

The ticket ID is a randomly generated number, expressed in base 64. In the case of the demo client the number is 168 bits long:

(demo-client.js:612). 
\begin{lstlisting}[frame=single,breaklines=true,basicstyle=\footnotesize]
itkt[I]=j=sMod.bin2b64(rndG.random(21)); //21*8=168 bits
\end{lstlisting}

This means the number of different ticket ids that could be generated are in the order of $10^{50}$, so it would be extremely unlikely for two different voters to generate the same ticket ID, assuming the algorithms they use are sufficiently random.
Once the ticket has been generated, it will be blinded with the opacity factor and sent to the SP for signing. 

The request will look like this:

\begin{verbatim}
param=ssign
token=dl0tim56ujtlu2pd2509cdm7j2
idsvy=845
opct=...blinded ticket...
\end{verbatim}

The SP will sign the ticket with the survey-specific key and return it to the client. It is very important for the signature to be survey specific because otherwise multiple participations could be allowed if a ticket generated for one survey is used to vote in another.

\newpage
The response from the SP looks like this:

\begin{verbatim}
<response>
  <signedTicket>...</signedTicket>
</response>
\end{verbatim}

The client now has the innermost ticket that he will send to the ballot box. 

\label{blindingDos}
If at this point the client is stopped before sending the ticket, he wont be able to vote anymore, because the SP will only sign a ticket once for every client.

This can be an issue. An attacker could deny a user the right to vote if he manages to keep the response from reaching the user. And if the users stops the process for some reason like a loss of connectivity, loss of power, or a crash in the client software, he will also be unable to vote.

To solve this issue, instead of marking the user and preventing him from getting another blindly signed ticket, the SP could allow a user to ask for a blind signature more than once, as long as the request is identical to his first request. On the client side, before sending a signature request, the ticket should be saved on disk so it can be restored and used again in case of problems on the client side (if a different blinded ticket is provided, the SP will refuse to sign it). 

Now that we have the inner ticket, it must be encrypted several times to be sent through the anonymity network, and finally the ballot box.

The innermost participation package will contain the ticket, its signature and the form contents. All of this is encrypted with the SP ballot box public key using RSA and RC4. The payload is encrypted with RC4, and stored in the \textit{cryp} field. The RC4 key is encrypted with rsa and store in the \textit{evkey} field. 

A package contains various fields:

\begin{itemize}
\item cryp : the encrypted payload
\item chk : an md5 checksum of the package
\item url : destination url of the next node
\item chal : the challenge expected from the next node
\item pChal : the challenge to be returned to previous node
\item sndD : timestamp when the package must be relayed
\item cloD : timestamp when the tallying will begin
\item rejD : timestamp when it's safe to discard the package in case it couldn't be delivered
\end{itemize}

\begin{figure}[!h]
\centerline{\includegraphics[width=1.0\textwidth]{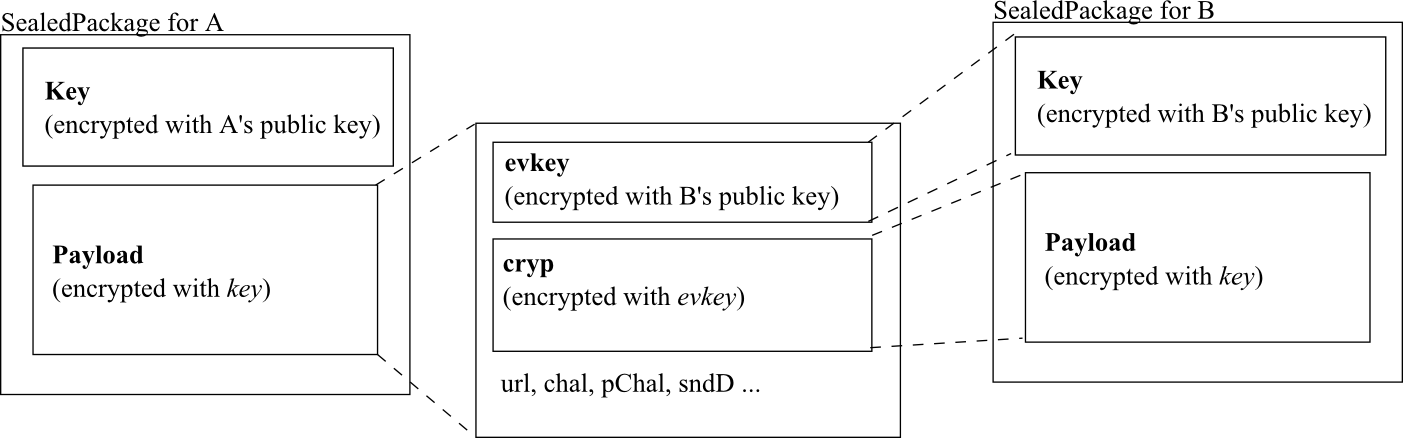}}
\caption{Package encapsulation.}
\label{fig:encapsulation}
\end{figure}

The package will then be recursively encrypted and encapsulated in an outer package as shown in \myfig{\ref{fig:encapsulation}}, once for each node in the route, starting from the last one. Every node will only be able to decrypt its own layer, reading the time the package must be relayed to the next node, the url of the next node, and the challenge that must be returned to the previous node. The sending node will delete a package from the queue only if the challenge received is the expected one, else it will try resending it.

\newpage
Let's see an example of the process with two nodes to make it clearer:

\begin{itemize}
\item The innermost package (A) is first created. It contains the signed ticket and the survey data, and it's encrypted with the ballot box public key.
\item The encrypted package A is encapsulated in a bigger package B, inside its \textit{cryp} field (see \myfig{\ref{fig:encapsulation}}). The url field will contain the url of the ballot box. \textit{Chal} will contain, for example, 999 and \textit{pChal} 888. The entire package is encrypted with the public key of the last node in the route (\textit{nodeL}).
\item This encrypted package is again encapsulated inside the \textit{cryp} of a bigger package C. The url field will contain the url of the last node (\textit{nodeL}). \textit{Chal} will be 888. The entire package is encrypted with the public key of the first node (\textit{nodeF}).
\end{itemize}

At the end of the process we have one big package with several layers of encryption. The package will be sent to the first node, \textit{nodeF}, with a request that looks like the one in \myfig{\ref{fig:request}}. \textit{NodeF} will use its private key to decrypt the "key" field, and use the decrypted symmetric key to decript the "payload" field (see \myfig{\ref{fig:encapsulation}}). In the paload it will find package C, from which it can read the url of the next node, \textit{nodeL}, and the expected challenge from that node (888). The fields \textit{cryp} and \textit{evkey} will be unreadable because they are encrypted with \textit{nodeL}'s public key. When the package must be relayed a new sealedPackage like the one in \myfig{\ref{fig:request}} will be created, using the \textit{evkey} field of package B for the "key" field, and the \textit{cryp} field of package B for the "payload" field (\myfig{\ref{fig:encapsulation}}). This sealedPackage will be sent to \textit{nodeL} which, using its private key, will be able to decrypt it and obtain package B (which was unreadable to \textit{nodeF}). The node will reply to \textit{nodeF} with the content of the field \textit{pChal} (888). It will also read the url field, and when \textit{sndD} is reached, relay the inner, encrypted package to the next node, which in this case is the ballot box.  

\subsubsection{Delivery}

\begin{figure}[!h]
\centerline{\includegraphics[width=0.7\textwidth]{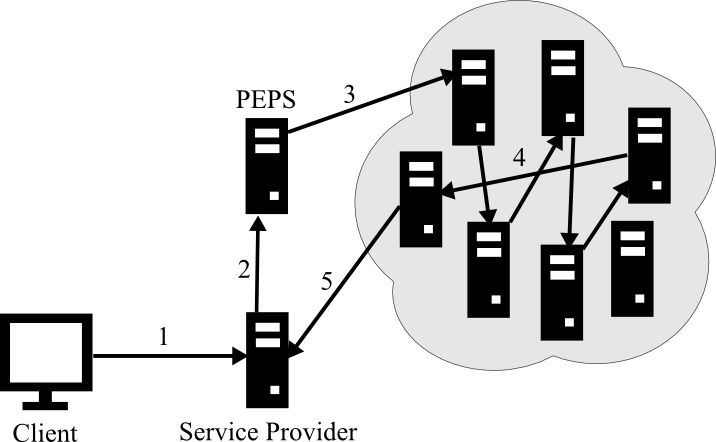}}
\caption{A bundle containing the packets is proxied to the PEPS (2), which relays it to the anonymity network (3). After bouncing between nodes a few times (4) the package will reach the ballot box on the SP (5)}
\label{fig:stage3}
\end{figure}

Once the packages for the various alternative routes are created they are proxied through the PEPS and sent to the network (\myfig{\ref{fig:stage3}}). The packages can all be put inside a single bundle (the demo client sends them in separate bundles though).

The bundle is sent to the SP with the request to proxy it to the S-PEPS:

\begin{verbatim}
param=proxy
token=12345
idsvy=12345
bundle=<bundle>...</bundle>
\end{verbatim}

The SP will apply an XMLDsig and send the package to the S-PEPS, which acts as the first node in the anonymity network.

Every node in the network will decrypt the package with its private key when its received, and add it to a local queue. Periodically the node will run the function \verb|processQueue()| (OutboundPackageHandler.java), and each package that needs to be sent (because the \textit{sndD} time has passed), will be sent to the relative node. 

The package will only be removed from the queue if a correct challenge response is received by the next node, otherwise it will be kept in the queue to retry at a later time. 

The packages are kept in the queue until the \textit{ReceivalDue} time + a \textit{rejectTime} variable which is set up in the configuration file (anonymity.properties). The default value is 24 hours (86400000ms). The \textit{ReceivalDue} value is taken from the \textit{CloD} value in the package.

\label{queueDos}
This kind of queue management is risky, because in the current implementation there's no defence against maliciously forged CloD values and challenge values. This could be exploited to force the nodes to continuously keep bouncing packets between each other. The queues could grow indefinitely causing memory starvation on a target node.

In particular a package can be created such as the expected challenge from the next node will be different from the challenge encrypted in the package destined to that node. This way the package will remain in the sender queue forever because the challenge response will always be wrong, and the sender will keep trying to send it until \textit{cloD} is reached. \textit{cloD} can also be set by the attacker to an arbitrary distant date, so the packages will be kept in the queue forever. 

It would also be hard to trace the source of the attack and block these malicious packets, as the challenge "bomb" can be hidden at any layer of encryption, making it hard to identify the original sender of the malicious packet. 

A malicious packet could be created in such a way:

\begin{itemize}
\item The inner ticket can be anything, even random bits (it doesn't matter because it will never reach the ballot box).
\item The \textit{challenge} field of layer N will be a random number
\item The \textit{expected challenge} field of layer N-1 will be a number different from the previous one, and the \textit{cloD} field will be an arbitrarily big value
\item All other layers will be fabricated normally
\end{itemize}

A package crafted this way is designed to remain forever stuck in the queue at node N-1. The first nodes will see it as a regular package, indistinguishable from the others, it will reach node N-1 from another arbitrary node in the network. By sending enough packages crafted this way, an attacker could deliberately congest node N-1.

A solution to this problem would be to discard a packet if the challenge response received is incorrect, instead of keeping it in the queue (OutboundPackageHandler:280).

The same could happen if on a certain layer the url of the next node is deliberately set to an unreachable address, and the cloD value is set to a high number. 

To prevent this from happening, instead of trying until cloD, a maximum delay from the sndD date or a maximum number of tries could be set.

Once the closing time is reached the Ballot Box server will receive all participations from the anonimity network. Given that the same participation will be sent through multiple alternative routes, the ballot box only considers the first ticket it receives with a certain ticket id. Subsequent tickets with the same ID, coming from alternative routes,  will be discarded.  

At the end of the election the SP should show publicly all the tickets received, so that users can check that their vote was counted. (only they will know which ticket ID belonged to them). If a voter sees that his vote wasn't included in the ballot box, currently there is no way to know which server was the one maliciously dropping his vote. If all servers were required to return a signed message to confirm the recieval of a package, then with their collaboration it would be possible to identify the server who maliciously dropped the vote.

\section{Experimental evaluation}

\subsection{Setting up virtual machines for sniffing}

For testing purposes the code was deployed on three virtual machines using VirtualBox, two running the PEPS software, and one acting as a demo Service Provider. 

In order to run the virtual machines on the same virtual network, and be able to sniff the traffic between them, for every virtual machine go to \verb|Device->Network Adapter| and choose attached to "Internal network" and "intnet" as the name of the network.

Then, on the guest OS (xubuntu 12.04) the machines have been set up to use a static IP address:

\begin{verbatim}
192.168.0.2  : node1 (a PEPS serer)
192.168.0.3  : node2 (another PEPS server)
192.168.0.99 : demoSP
\end{verbatim}

To sniff the packets we used VirtualBox's network tracing feature. It can be enabled with this command, before running the VM:

"VBoxManage modifyvm [vm-name] --nictrace[adapter-number] on --nictracefile[adapter-number] file.pcap"

For example:

"VBoxManage modifyvm "PEPS" --nictrace1 on --nictracefile1 file.pcap"

\subsection{PEPS installation}

The PEPS software was installed and deployed under Tomcat 7.

Before compiling and installing the PEPS it is necessary to set up the correct keystores and edit configuration files. 

Other than the files listed in the documentation, it is also required to modify the files in \verb|PEPS\src\main\config\embedded\SignModule_*****.xml| with the correct certificate information (serial number and issuer).

The logs, useful for debugging, can be found inside the \verb|webapps| subfolder in the \verb|CATALINA_HOME| directory, usually:

\verb|C:\Program Files (x86)\Apache Software Foundation\Tomcat 7.0| (on Windows) 

or \verb|/var/lib/tomcat7/| (on Linux)

\subsubsection{Anonymity module installation}

The anonymity module can be installed as described in the official guide, but some extra steps are necessary.

In the configuration file \verb|anonymity.properties| two extra lines were nrequired:

\begin{verbatim}
#Anonymity properties
anonymity.keystore.path=file:/pathToKeystore.jks
anonymity.keystore.pass=password
\end{verbatim}

The aliases in the example keystore should also be changed:

\begin{verbatim}
From:     To:
node1 --> node1
node2 --> node2.current
node3 --> node3.current
\end{verbatim}

The software expects a ".current" suffix for all aliases except the node it's running on. This is used to deal with the update of certificates.

The Anonymity module also needs the ojdbc6 library (version 11.2.0.2.0) which, if not present, should be downloaded directly from Oracle's website (\url{http://www.oracle.com/technetwork/database/enterprise-edition/jdbc-112010-090769.html}) , after accepting their license.

The ojdbc jar can be manually added to the maven repositoy with the command:

\begin{verbatim}
mvn install:install-file -Dfile=./ojdbc6.jar -DgroupId=com.oracle \
-DartifactId=ojdbc6 -Dversion=11.2.0.2.0 -Dpackaging=jar
\end{verbatim}

When installing the database, make sure that the \verb|addres| value of nodes, corresponds to the real, absolute addresses of test nodes.

Once the module has been compiled, the PEPS code must be compiled again, with the anonymity option enabled:

\verb|mvn clean package -P embedded -P anonymity -DskipTests|

\subsection{Anonimity DemoSP installation}

For testing purposes the demoSP was installed inside a virtual machine running Ubuntu 12.04.

Apache was used as a web server (with PHP support), and MySQL for the database functionality. 

The database schema to be installed can be found in \verb|Anonymity_demoSP/db.sql|, it is used for storing the information about surveys and storing the tickets received by the ballot box. 

Before running the demoSP, the configuration file, \verb|common.php| must be edited with the correct information.

In particular the mysql credentials must be changed, as well as the url of the S-PEPS. 

For the PEPS to accept bundles signed by the demo SP, the SP public key must be imported in the PEPS keystore. This can be done with the command:

\verb|keytool -import -alias demoSP -keystore storkNodeKeys.jks -file demoSP_cert.pem|

\begin{figure}
\centerline{\includegraphics{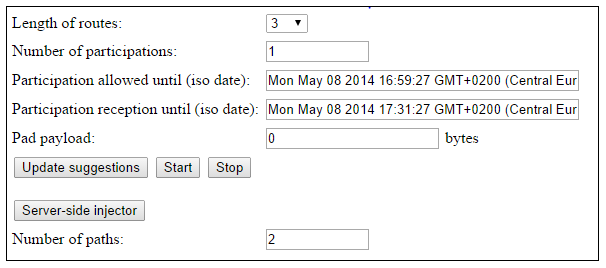}}
\caption{Settings shown on index.html}
\label{fig:settings}
\end{figure}

On the index.html page there are some settings for testing, as shown in \myfig{\ref{fig:settings}}: 

\begin{itemize}
\item \textbf{Participations allowed until} corresponds to the \textit{cloD} field.
\item \textbf{Participations reception until} corresponds to the \textit{endD} field.
\item \textbf{Number of paths} is the number of alternative routes to be used for a single ticket.
\item \textbf{Update suggestions} changes the dates according to the current date.
\item \textbf{Server-side injector} is used to generate partecipations directly on the SP and inject them in the network (survey.php:129). 
\end{itemize}

\label{padding}
The payload padding can be important in some cases. Consider a survey where option B is 50 bytes longer than option A, it would be easy for any entity in the network to differentiate between tickets voting for A and tickets voting for B, just by looking at the size of the encrypted package. 

It would be important, therefore, to pad packets with a random length, to make it harder to differentiate encrypted votes. Even better than using a big random length pad, one could generate a padding such that all options in the survey have the same length. Otherwise malicious nodes could calculate, given a certain package length, how likely it is to be a vote for candidate A or candidate B, and drop the packet according to this information. 

Even if there isn't a 100\% certainty, on a big enough number of packets this could skew the survey towards a certain option. In particular the SP or the PEPS could do this. Nodes in the network can drop packets maliciously as well, but they would need to control nodes on all the alternatives routes to drop packets successfully. 

It is also important to notice that, while the SP doesn't know who is voting for what, in this configuration it knows with certainty who voted and who chose to abstain. To keep this information private, an extra option should be present in the survey to account for abstainees, generating a package indistinguishable from a regular vote. 

\section{Comparison with TOR}

The onion routing strategy used to anonymize users is very similar to the one used in TOR (The Onion Router), with some minor differences.

\subsection{Encryption}

\label{encryption}
For the stream encryption TOR uses 128-bit AES in counter mode, with an IV of all 0 bytes\cite{tor}, and for the public-key cipher, it uses RSA with 1024-bit keys and a fixed exponent of 65537. Although recent versions have increased the key size, most users still run older versions with 1024 bit keys.

This PEPS implementation also uses RSA keys of 1024 bit, although the key length can be increased in the configuration file: 

(common.php:47)
\begin{lstlisting}[frame=single,breaklines=true,basicstyle=\footnotesize]
$keyLength='1024';
\end{lstlisting}

For the stream cypher, RC4 is used, with a key length of 128 bit.

(demo-client.js:1793)
\begin{lstlisting}[frame=single,breaklines=true,basicstyle=\footnotesize]
var S=rndG.random(16); //16*8=128 bits
\end{lstlisting}

RC4 is subject to statistical attacks but this should not be an issue in this use-case, as it is only used as a one-time pad. 

The encryption strength of both systems is comparable. It would be advisable to switch from 1024 bit keys to bigger keys, as 1024 bit keys are considered creckable by government entities as of 2014. The NIST suggests a key size of 2048 bits or higher. 

\subsection{Bootstrapping}
\label{boot}

The way the client becomes aware of nodes in the network is very important for an anonymizing protocol such as these. The TOR and PEPS implementations are different in this regard.

The PEPS approach is very centralized. The list of nodes is retrieved from the PEPS through https. This means the user must trust the entity running the PEPS, and its certificate authority, or he could be served a malicious list of nodes. 

TOR has a more distributed approach. There are a number of known, independent "Directory Authorities", which are contacted by the clients to learn about the nodes in the network. There are 10 official Directory Authorities, and they are hardcoded in the TOR client. Alternative authorities can be used, but this is advised against to avoid splitting of the network. 

Nodes in the network (\textit{routers}, in TOR's terminology) regularly contact the Directory Authorities to advertise their status. Every day, starting at 00:00 am, the authorities vote to reach a new consensus on the status of the network. Routers are considered active only if the majority of authorities has included them in their "vote". A \textit{consensus document} on the status of the network is thus generated. The users download the consensus document at a random time from a random Directory Authority. The actual protocol is more complicated than this, for details see\cite{torAuthorities}.

In this case, the end user only needs to trust the majority of the Directory Authorities. If a single authority is compromised, it is not an issue. 

The approach chosen by the STORK project is simpler, and has a single point of failure, but it shouldn't be a big issue in this case. In fact, even if the attacker manages to serve a compromised list of nodes, the votes will still remain anonymous as long as the SP is not also compromised. What an attacker could do by serving a compromised list of nodes, is denying a user the right to vote. Given that the list of nodes is passed from the PEPS to the client directly, the PEPS will know the IP addresses of voters requesting the node list, and could make it impossible for some users to vote, based on their IP address.

If the PEPS cannot be trusted to be impartial in the delivery of node lists, it would be better to use an approach more similar to TOR's, with multiple entities serving the public node list, so everyone knows they're being served the same list, unless all servers are colluding.  

\subsection{Timing attacks}

This is where the two systems differ the most. TOR was created mainly to anonymize access to websites, or general purpose connections. It's therefore important to have a latency which is not too high, to keep the services usable, while making the users hard to identify. In order to do this, TOR needs a lot of nodes and a lot of traffic going between them, to make time-based traffic analysis harder.

Our use case is different. Using tor wouldnt be sufficient to anonymize users in the case of an election, because their votes need to be authenticated, so even if their IP is hidden, the votes will still be tied to their accounts. Even if a blind signature scheme is used, it would still be possible to trace voters with some degree of certainty, based on the time when their vote reaches the ballot box. 

What the STORK anonymity network does to solve this issue, other than hiding the users IP, is also making sure all the packets (the votes) reach the server within a specific time window, and in a completely random order, making it impossible for the server to link packets with users. This cannot be done in the case of TOR, which was engineered to be used in many different situations, where additional, artificial delays are not desirable.

This also means that the anonymity network in STORK doesn't need as many nodes, or as much traffic, because it is resilient to timing analysis attacks by design. Increasing the number of nodes is mainly useful to increase availability and reduce the risk that all nodes are controlled by a malicious entity. This is still a risk in any case, though, given that the list of nodes is received by a single entity, no matter how many the nodes in the network.

\section{Conclusions}

This anonymity protocol can be useful in many cases, and the reference implementation works, but there are some security concerns which should be addressed before deployment: 

\begin{itemize}
\item the client should use a better entropy buffer for the generation of the \textit{opacity factor}, otherwise the SP could easily brute force it, and unblind a blinded ticket (see \ref{entropyWeakness})
\item the \textit{cloD} value (the time when the votes are tallied) must be verifiably the same for all voters (see \ref{cloDattack}), otherwise voters can be deanonymized based on the time their vote is received
\item the client should be allowed to request a blind signature more than once, as long as the blinded ticket provided is the same. Otherwise, connectivity problems (malicious or involuntary) could prevent a user from voting (see \ref{blindingDos})
\item nodes should discard packets if the challenge received from the next node is incorrect, and if the next node is offline, only keep trying for a limited number of times. Otherwise, queues can grow indefinitely and cause memory starvation on a target node (see \ref{queueDos})
\item an option to abstain should be present in every survey, and all options should have the same length in bytes, or a long, random padding should be used. This makes it impossible to profile encrypted votes based on the package length (see \ref{padding})
\item longer RSA keys of 2048 bits or more would be preferred, to protect against targeted attacks from government entities (see \ref{encryption})
\item it would be better if the list of nodes wasn't downloaded from a single entity (the PEPS). A distributed approach like the one used in TOR could be preferred (see \ref{boot})
\end{itemize}

As it stands, if the SP and the PEPS are controlled by the same entity, or they are colluding to deanonymize users, they can easily do so. Therefore a system like this should not be trusted to grant anonymity in the case of a government-run election.  

The reference implementation has been tested and it works for a simple test case, but further tests would be required to observe the behaviour under heavy load on a deployment server.


\newpage
\begin{thebibliography}{99}
%
%


\bibitem{stork1}
Herbert Leitold, 
``Challenges of eID Interoperability: The STORK Project'', 
Privacy and Identity Management for Life,
Volume 352,  2011,
pp. 144-150

\bibitem{stork2}
V. Koulolias et al.
``STORK e-privacy and security'',
5th International Conference on Network and System Security, 2011,
pp. 234 - 238 

\bibitem{chaum}
Chaum, David, 
''Blind signatures for untraceable payments'',
Advances in Cryptology Proceedings of Crypto 82 (3), 1989, pp. 199-203

\bibitem{chaumsec}
M. Bellare and C. Namprempre and D. Pointcheval and M. Semanko,
"The One-More-RSA-Inversion Problems and the Security of Chaum's Blind Signature Scheme"
Journal of Cryptology (16),
2003, pp. 185-215

\bibitem{storkSource}
Stork 2.0 source code, 
\url{https://www.eid-stork2.eu/index.php?option=com_content&view=article&id=61}

\bibitem{distributed}
\url{http://blogs.distributed.net/2002/09/25/00/00/bovine/}

\bibitem{random}
Amit Klein,
"Temporary user tracking in major browsers and Cross-domain information leakage and attacks",
2008, chapter 2

\bibitem{fortuna}
\url{https://clipperz.is/security_privacy/crypto_algorithms/#prng}

\bibitem{tor}
Roger Dingledine, Nick Mathewson,
"Tor Protocol Specification", 
\url{https://gitweb.torproject.org/torspec.git?a=blob_plain;hb=HEAD;f=tor-spec.txt}

\bibitem{torAuthorities}
"Tor directory protocol, version 3", 
\url{https://gitweb.torproject.org/torspec.git?a=blob_plain;hb=HEAD;f=dir-spec.txt}

\end{thebibliography}
\end{document}